\documentclass[twoside,leqno, twocolumn]{article}

\usepackage[letterpaper]{geometry}

\usepackage{float}
\usepackage{ltexpprt}
\usepackage{hyperref}

\usepackage{amsmath,bm}
\usepackage{tikz}
\usetikzlibrary{calc}
\usepackage{pgfplots}

\usepackage{listings}

\definecolor{clr-background}{RGB}{255,255,255}
\definecolor{clr-text}{RGB}{0,0,0}
\definecolor{clr-string}{RGB}{163,21,21}
\definecolor{clr-namespace}{RGB}{0,0,0}
\definecolor{clr-preprocessor}{RGB}{255,128,0}
\definecolor{clr-preprocessor2}{RGB}{153,76,0}
\definecolor{clr-preprocessor3}{RGB}{102,0,204}
\definecolor{clr-keyword}{RGB}{0,0,255}
\definecolor{clr-type}{RGB}{43,145,175}
\definecolor{clr-variable}{RGB}{0,0,0}
\definecolor{clr-constant}{RGB}{111,0,138} 
\definecolor{clr-comment}{RGB}{0,128,0}

\definecolor{c1}{rgb}{0,0.4470,0.7410}
\definecolor{c2}{rgb}{0.8500,0.3250,0.0980}
\definecolor{c3}{rgb}{0.9290,0.6940,0.1250}
\definecolor{c4}{rgb}{0.4940,0.1840,0.5560}
\definecolor{c5}{rgb}{0.4660,0.6740,0.1880}
\definecolor{c6}{rgb}{0.3010,0.7450,0.9330}
\definecolor{c7}{rgb}{0.6350,0.0780,0.1840}
\definecolor{c8}{rgb}{0,0.4325,0.1746}

\usepackage[T1]{fontenc}

\definecolor{light-gray}{gray}{0.95}

\lstdefinestyle{cstyle}{
    language=C++,
	backgroundcolor=\color{light-gray},
	basicstyle=\scriptsize\fontfamily{ptm}\color{clr-text}, 
	stringstyle=\color{clr-string},
	commentstyle=\color{clr-comment},
    keywordstyle = {\color{clr-type}},
    keywordstyle = [2]{\color{clr-preprocessor3}},
    morekeywords = [2]{Kokkos, atomic_add, kokkos_builtin_derivative, parallel_sum, deep_copy, View, LayoutLeft, LayoutRight, LayoutStride, parallel_for, parallel_reduce, TeamPolicy,KOKKOS_LAMBDA, KOKKOS_INLINE_FUNCTION,TeamThreadRange,ThreadVectorRange,TeamVectorRange,Sacado,Rad,scratch_memory_space},
	tabsize=4,
	captionpos=b,
    columns=fullflexible,
	basewidth={0.6em,0.55em},
	moredelim=[is][\color{clr-preprocessor}]{|>}{<|}	
}

\tikzset{
  mycircle/.style={
    draw, circle, c1, thick, minimum height=0.35cm, minimum width=0.35cm
  }
}

\tikzset{
  mycylinder/.style={
    draw, cylinder, rotate=210, c1, thick, minimum height=0.5cm, minimum width=0.35cm, aspect=1.0
  }
}

\pgfplotsset{compat=1.18} 

\begin{document}

\newcommand\relatedversion{}
\newcommand{\codestyle}[1]{\lstinline[style=cstyle,basicstyle=\small\ttfamily]|#1|}

\title{\Large Performance Portable Gradient Computations Using Source Transformation\relatedversion}
\author{Kim Liegeois\thanks{Sandia National Laboratories.}
  \and Brian Kelley\footnotemark[1]
  \and Eric Phipps\footnotemark[1]
  \and Sivasankaran Rajamanickam\footnotemark[1]
  \and Vassil Vassilev\thanks{Princeton University.}}

\date{}

\maketitle


\fancyfoot[R]{\scriptsize{Copyright \textcopyright\ 20XX by SIAM\\
    Unauthorized reproduction of this article is prohibited}}





\begin{abstract} \small\baselineskip=9pt

  Derivative computation is a key component of optimization, sensitivity analysis, uncertainty quantification, and nonlinear solvers.
  Automatic differentiation (AD) is a powerful technique for evaluating such derivatives, and
  in recent years, has been integrated into programming environments such as Jax, PyTorch, and TensorFlow to support derivative computations
  needed for training of machine learning models, resulting in widespread use of these technologies.
  The C++ language has become the de facto standard for scientific computing due to numerous factors,
  yet language complexity has made the adoption of AD technologies for C++ difficult,
  hampering the incorporation of powerful differentiable programming approaches into C++ scientific simulations.
  This is exacerbated by the increasing emergence of architectures such as GPUs, which have limited memory capabilities and require massive thread-level concurrency. Portable scientific codes rely on domain specific programming models such as Kokkos \cite{trott2021kokkos} making AD for such codes even more complex.

  In this paper, we will investigate source transformation-based automatic differentiation using Clad to automatically generate portable
  and efficient gradient computations of Kokkos-based code. We discuss the modifications of Clad required to differentiate Kokkos abstractions.
  We will illustrate the feasibility of our proposed strategy by comparing the wall-clock time of the generated gradient code with the wall-clock time of the
  input function on different cutting edge GPU architectures such as NVIDIA H100, AMD MI250x, and Intel Ponte Vecchio GPU.
  For these three architectures and for the considered example, evaluating up to $10\,000$ entries of the gradient only took up to $2.17\times$ the wall-clock time of evaluating the input function.

\end{abstract}

\section{Introduction}
Derivative computations are critical for many areas of scientific computing such as optimization, sensitivity analysis, uncertainty quantification, and nonlinear solvers.
Automatic differentiation (AD) is a powerful technique for evaluating such derivatives~\cite{griewank_automatic_1991, berz_computational_1996, corliss_automatic_2002, bucker_automatic_2006, griewank_evaluating_2008, bischof_advances_2008, forth_recent_2012, Naumann2012TAo}.
AD has recently seen widespread usage for training machine learning models in Python frameworks like Jax~\cite{jax2018github,frostig2018compiling}, PyTorch~\cite{paszke2017automatic,NEURIPS2019_9015}, and TensorFlow ~\cite{tensorflow_developers_2022_6574269,tensorflow2015-whitepaper}.
C++ continues to be the de facto language for scientific computing due to numerous factors, but the complexity of the language makes AD difficult.
This has hampered the incorporation of powerful differentiable programming approaches~\cite{JMLR:v18:17-468} into C++ scientific codes.
The emergence of GPUs as general-purpose accelerators has further increased this challenge, since GPUs have a different memory management paradigm and require massive thread-level concurrency.

There are two main implementation paradigms used for AD, operator overloading and source transformation.
The former is simpler to support in general-purpose C++ code, but less efficient.
The latter provides much better opportunities for generation of optimized derivative code, but is difficult in general C++ code and has been more successful in simpler languages such as C and Fortran. Numerous tools for both approaches are available for a variety of programming environments, with a comprehensive listing available at \url{www.autodiff.org}.  There are also two primary methods of propagating derivatives through computations considered in AD, forward and reverse mode.  Forward mode is useful for propagating derivatives with respect to small numbers of independent variables, and typically is used for directional derivatives and square Jacobian computations.  Reverse mode is very efficient for computing derivatives of a small number of quantities with respect to a large number of independent variables.  This is the case during the training of machine learning/deep learning models
with large numbers of weights or when computing the gradient of an objective function of a PDE constrained optimization problem with a large number of parameters.

For relevance to scientific computing, AD techniques must be applicable to parallel computations and codes with control flow operations.  Accordingly, AD tools and techniques have been developed for differentiating codes relying on parallel programming environments such as distributed message passing with MPI~\cite{Hovland:1998,Faure:1999,Carle2002ADM,Utke2009TAM,Pascual2012NHo} and multithreading through OpenMP~\cite{Bucker2001BTA,Bucker2002ELS,Bucker:2004,Bucker:2008,Bischof2008PRM,Letschert2012ESi,Forster:2014,Huckelheim:2022}.  With the emergence of exascale architectures, however, tools must be developed that support differentiation of code executing on GPUs.  In prior work, we have demonstrated good performance for operator overloading-based forward mode AD on GPU architectures~\cite{phipps2022AD} through the Sacado AD package ~\cite{SacadoURL,Phipps2008LST,Phipps2012LST}
and its integration with the Kokkos performance portability library~\cite{trott2021kokkos} for shared-memory parallelism on CPUs and GPUs.
However, while the Clad~\cite{vassilev2015clad}, Enzyme~\cite{enzyme}, and LAGrad~\cite{peng2023lagrad} tools provide some reverse mode capabilities along with limited support for GPUs, there are no well-established approaches for reverse-mode
differentiation of general C++ code on GPUs. This is due to challenges in generating efficient derivative code while preserving performance portability of the original code~\cite{huckelheim2022source, huckelheim2018parallelizable, revels2018dynamic, rootfitting, NEURIPS2020_9332c513, enzyme}. Given the importance of reverse mode for optimization and machine learning, it is critical for AD tools to produce efficient derivative code for extreme-scale scientific simulations that execute on modern HPC hardware. Furthermore, the increasing diversity of CPU and GPU architectures can be mitigated by performance-portability libraries such as Kokkos.  AD support for Kokkos-based codes is therefore valuable for many large-scale scientific workflows.

Kokkos consists of high-level programming idioms implemented using C++ templates.  These include generic parallel execution patterns such as for-each, reduction, and scan.
Performance portability is achieved through tightly integrated mappings of loop bodies, parallel iteration spaces, and multidimensional arrays onto the target architecture's parallel programming model and hardware.

Building on top of Kokkos, Sacado provides general, performance-portable forward mode AD. Special scalar types support the usual C++ arithmetic operators, but represent primal and derivative values together.
Partial specializations of the Kokkos multidimensional array (\lstinline[style=cstyle,basicstyle=\small\ttfamily]|Kokkos::View|) for these types add a hidden dimension for efficiently storing the derivative data ~\cite{phipps2022AD}.
Sacado and its Kokkos integration provide high-performance derivative capabilities to numerous Office of Science and NNSA extreme scale applications, including Albany for solid mechanics and land ice modeling \cite{Salinger2016,MPASAlbany2018}, Charon for semiconductor device modeling \cite{CharonUsersManual2020} and multiphase chemically reacting flows \cite{Musson2009}, Drekar for computational fluid dynamics (CFD) \cite{Sondak2021,Shadid2016}, magnetohydrodynamics \cite{Shadid2016mhd} and plasma physics \cite{Crockatt2022,Miller2019}, Xyce for electronic circuit simulation~\cite{xyceTrilinos,xycePCE}, and SPARC for hypersonic fluid flows~\cite{SparcValidation}.

In this paper, we investigate the automatic generation of Kokkos-based gradient code using source-to-source transformation.
In particular, we developed a prototype based on Clad~\cite{vassilev2015clad} that adds support for core Kokkos features and shows that automatically generating Kokkos-based gradient code is feasible.
In Section~\ref{sec::Background}, we provide a brief description of Clad~\cite{vassilev2015clad} and contrast it with other LLVM-based tools such as Enzyme~\cite{enzyme} and LAGrad~\cite{peng2023lagrad}. Then we explain why none of those tools were
able to generate Kokkos C++ code for computing a gradient.
In Section~\ref{sec::Methodology}, we describe the differentiation rules for Kokkos features that we implemented in Clad.
In Section~\ref{sec::Results}, we apply the Kokkos-aware Clad to generate the gradient code of an objective function and we evaluate the performance of the gradient evaluation relative to the objective function evaluation.
Finally, in Section~\ref{sec::Conclusion}, we summarize our contributions and propose several directions for future work.

\section{Background}
\label{sec::Background}

LLVM ~\cite{lattner2004llvm} is an open source compiler framework. The LLVM infrastructure provides production quality programming language building blocks for analysis and optimization. Its applications span vectorization~\cite{moll2019multi}, automatic parallization~\cite{kalms2018automatic}, language interoperability~\cite{binder}, and AD~\cite{vassilev2015clad,enzyme,peng2023lagrad}.



\begin{figure}[htbp]
  \begin{center}
    \begin{tikzpicture}[node distance=1cm]
      \node (Input) {Input file};
      \node (AST) [below of=Input] {AST};
      \node (MLIR) [below of=AST] {MLIR};
      \node (llvmIR) [below of=MLIR] {LLVM-IR};

      \node (Output) [right of=Input, node distance=3cm] {Output file};
      \node (pAST) [right of=AST, node distance=3cm] {$\partial$AST};
      \node (pMLIR) [right of=MLIR, node distance=3cm] {$\partial$MLIR};
      \node (pllvmIR) [right of=llvmIR, node distance=3cm] {$\partial$LLVM-IR};
      \node (pbinary) [below of=pllvmIR] {$\partial$binary};

      \draw[ultra thick, c1] ($(Input.south)-(1em,0)$) -- ($(AST.north)-(1em,0)$);
      \draw[ultra thick, c1] ($(AST.south)-(1em,0)$) -- ($(MLIR.north)-(1em,0)$);
      \draw[->, ultra thick, c1] ($(MLIR.south)-(1em,0)$) -- ($(llvmIR.north)-(1em,0)$);

      \draw[->, ultra thick, c1] (llvmIR) -- (pllvmIR);
      \draw[->, ultra thick, c1] ($(pllvmIR.south)-(1em,0)$) -- ($(pbinary.north)-(1em,0)$);

      \draw[->, dashed, ultra thick, c2] (Input) -- (AST) -- (MLIR);
      \draw[->, ultra thick, c2] (MLIR) -- (pMLIR);
      \draw[->, ultra thick, c2] (pMLIR) -- (pllvmIR) -- (pbinary);

      \draw[->, ultra thick, c3] ($(Input.south)+(1em,0)$) -- ($(AST.north)+(1em,0)$);
      \draw[->, ultra thick, c3] (AST) -- (pAST);
      \draw[<-, ultra thick, c3] ($(Output.south)+(1em,0)$) -- ($(pAST.north)+(1em,0)$);
      \draw[ultra thick, c3] ($(pAST.south)+(1em,0)$) -- ($(pMLIR.north)+(1em,0)$);
      \draw[ultra thick, c3] ($(pMLIR.south)+(1em,0)$) -- ($(pllvmIR.north)+(1em,0)$);
      \draw[->, ultra thick, c3] ($(pllvmIR.south)+(1em,0)$) -- ($(pbinary.north)+(1em,0)$);
    \end{tikzpicture}
  \end{center}
  \caption{Illustration of the different level of abstraction of the LLVM stack and where \textcolor{c1}{\textbf{Enzyme}}, \textcolor{c2}{\textbf{LAGrad}}, and \textcolor{c3}{\textbf{Clad}} perform the differentiation.
    A dashed line is used between the input file and MLIR for LAGrad because LAGrad needs a MLIR file as an input. Therefore, in order to use it with a C++ file, the file needs to be first lowered to a MLIR representation and this is done with other tools such as Polygeist~\cite{moses2021polygeist}.}
  \label{differentiation_level}
\end{figure}
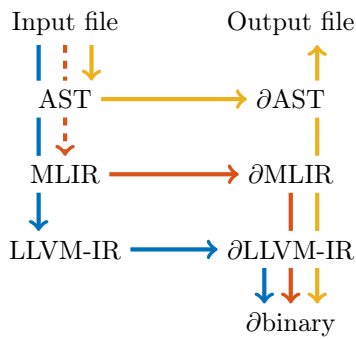

Clad~\cite{vassilev2015clad}, Enzyme~\cite{enzyme}, and LAGrad~\cite{peng2023lagrad} implement AD  at different levels of abstraction. They all run ahead of execution, analyze the program representation and produce new code following the common AD rules.
Enzyme operates on \textit{LLVM IR}, an assembly-like intermediate representation. LAGrad operates on \textit{MLIR}, a generalization of LLVM IR which contains higher level program information. Finally, Clad operates on the close-to-source \textit{C++ AST}. The three abstraction levels are illustrated in Fig.~\ref{differentiation_level}.

Each abstraction level enables different AD functionalities:
\begin{itemize}
  \item LLVM IR is a low-level representation, which simplifies cross-language interoperability but loses high-level structure and type information from the source language. LLVM IR enables Enzyme to perform AD across multiple  frontends supported by LLVM, such as C, C++ and Julia. A benefit of LLVM IR is that optimization passes can be applied before performing AD. However, differentiation can be sensitive to certain passes. For example, loop unrolling might make the LLVM IR non-differentiable.
  \item MLIR preserves more structure, but it is still immature and currently no major compiler produces high-quality MLIR. MLIR also enables LAGrad to perform AD after optimization. MLIR input files can be generated from C/C++ source using Polygeist~\cite{moses2021polygeist}, however many C++ template idioms do not produce code and run only in the frontend.
  \item The AST preserves complete type information and access to the C++ compiler frontend at the cost of more difficult implementation of the AD functionality. The AST gives Clad full access to the C++ type system and templates. Clad can generate output in both source code and binary form, but there are fewer opportunities for optimization before differentiation.  However, Clad can use Clang's semantic analysis to perform its own optimizations.
\end{itemize}
While Clad is the only tool of the three that can directly generate source code, it is also feasible to emit Kokkos/C++ code from MLIR \cite{kelley2022unified}
and some tools are able to generate C++ from LLVM-IR~\cite{kalms2018automatic}.

Originally, none of these three tools were able to differentiate Kokkos/C++ code and produce Kokkos/C++ as output.
Clad (without our modifications) cannot differentiate AST nodes related to Kokkos features.
LAGrad cannot be used because there is currently no tool that supports lowering Kokkos-based C++ code to MLIR.
Enzyme cannot be used because there is currently no tool that supports raising LLVM-IR to Kokkos C++.

In order to understand the feasibility and performance of the automatic generation of performance portable gradient computations,
we decided to work at the AST level using Clad for two reasons.
First, gradient source code generation allows us to apply AD once and then compile and run the gradient on any architecture that Kokkos supports.
Second, the AST representation of the input source code maintains all high-level information about how Kokkos features are used. This makes it easier to write differentiation rules specific to Kokkos data types and functions.

\section{Methodology}
\label{sec::Methodology}

Our work had the goal of showing that source-to-source transformation is a feasible method for AD of Kokkos-based code.
For this purpose we restricted ourselves to a core subset of Kokkos features: \codestyle{View}, \codestyle{deep\_copy}, and \codestyle{parallel\_for}. The implementation of this work consists of explicit differentiation rules in Clad for these features. To discuss these rules, we use the notion of a computational graph \cite{bauer1974computational}.

This section only discusses the generation of gradient code using the reverse mode.
In order to evaluate a gradient with the reverse mode, the generated code typically has two passes: the forward pass is a duplicate of the computational graph of the input function, and the reverse pass traverses the same graph in reverse order. Code to compute the reverse pass is simply appended after the forward pass.

\subsection{Kokkos::View}

The first key Kokkos feature used in this paper is the \codestyle{View}, an abstraction for arrays of zero or more dimensions.
This \codestyle{View} is templated for the type of the elements, the number of dimensions, the memory layout, and the device type.
The size of each dimension of a \codestyle{View} can be either a compile-time constant or a runtime value (see Fig. \ref{fig::view_construction}).
There are three choices for the layout:
\codestyle{LayoutLeft} is column-major (like Fortran), \codestyle{LayoutRight} is row-major (like C/C++), and \codestyle{LayoutStride} allows arbitrary strides for each dimension.
The device type combines the ``execution space'' where code using the view should execute (e.g. OpenMP, CUDA) and the ``memory space'' where the data of the \codestyle{View} is allocated (host CPU or GPU).
The class provides member functions and data to access information on how the array is stored such as the rank (number of dimensions), the size of each dimension using \codestyle{extent}, and the stride associated with each dimension.

When a \codestyle{View} is constructed inside a function that the user wants to differentiate, Clad must generate a second \codestyle{View} of same data type, same memory space, and same size to store the values of the derivative associated to the \codestyle{View}.
In this work, we restricted ourselves to cases where the number of dimensions and the dimensions themselves are constant with respect to the derivative computations. In other words, the dimensions can still be defined at compile or runtime but a runtime dimension cannot depend on any active variable. Therefore, any accesses of data layout information should not be differentiated.
Those two first differentiation rules are illustrated in Fig.~\ref{fig::view_construction} where a \codestyle{View} named  \codestyle{y} is constructed. The  \codestyle{View} stores an array of type \codestyle{double}, uses the default layout and memory space, and has two dimensions: a runtime one and one known at the compilation time denoted with \codestyle{*} and \codestyle{[5]} respectively.
The \codestyle{View} constructor takes a name as argument and one value for each runtime dimension.
In this case, the first dimension is set to be the same as the first dimension of a different view \codestyle{x} accessed by the function call \codestyle{x.extent(0)}.
The generated code illustrated in the right part of Fig.~\ref{fig::view_construction} allocates two views: \codestyle{y} is the same as in the input code, and \codestyle{\_d\_y} (with the same type and dimensions) is added to store the derivative values.

\begin{figure}
  \begin{lstlisting}[style=cstyle]
Kokkos::View<double*[5]> y("y", x.extent(0));
    \end{lstlisting}
  \begin{center}
    $\Downarrow$
  \end{center}
  \begin{lstlisting}[style=cstyle]
Kokkos::View<double*[5]> _d_y("_d_y", x.extent(0));
Kokkos::View<double*[5]> y("y", x.extent(0));
    \end{lstlisting}
  \caption{Differentiation of a \codestyle{Kokkos::View} construction.}
  \label{fig::view_construction}
\end{figure}

Once views are constructed, the user can access their entries using the \codestyle{operator()}.
We assume that the indices used to access a view should not be differentiated. In the reverse pass, any read becomes a write and any write becomes a read as illustrated in Fig.~\ref{fig::view_accesses}.

\begin{figure}
  \begin{lstlisting}[style=cstyle]
y(j1 + 1) = 2.6 * x(j1);
    \end{lstlisting}
  \begin{center}
    $\Downarrow$
  \end{center}
  \begin{lstlisting}[style=cstyle]
y(j1 + 1) = 2.6 * x(j1);

//...

double _r_d1 = _d_y(j1 + 1);
double _r3 = 2.6 * _r_d1;
_d_x(j1) += _r3;
_d_y(j1 + 1) -= _r_d1;
    \end{lstlisting}
  \caption{Differentiation of \lstinline[style=cstyle,basicstyle=\small\ttfamily]|Kokkos::View| accesses.}
  \label{fig::view_accesses}
\end{figure}

\subsection{Kokkos::deep\_copy}

\begin{figure}
  \begin{center}
    \begin{tikzpicture}[node distance=0.5cm, scale=0.8, every node/.style={scale=0.8}]
      \begin{scope}[xshift=0cm]
        \node (n02) [mycircle] {};
        \node (n01) [mycircle, above of=n02] {};
        \node (n00) [mycircle, above of=n01] {};
        \node (n03) [mycircle, below of=n02] {};
        \node (n04) [mycircle, below of=n03] {};

        \node (n10) [mycircle, right of=n00, node distance=1cm] {};
        \node (n11) [mycircle, below of=n10] {};
        \node (n12) [mycircle, below of=n11] {};
        \node (n13) [mycircle, below of=n12] {};
        \node (n14) [mycircle, below of=n13] {};

        \draw[->, ultra thick, c1] (n00) -- (n10);
        \draw[->, ultra thick, c1] (n01) -- (n11);
        \draw[->, ultra thick, c1] (n02) -- (n12);
        \draw[->, ultra thick, c1] (n03) -- (n13);
        \draw[->, ultra thick, c1] (n04) -- (n14);
      \end{scope}
      \begin{scope}[xshift=2cm]
        \node (n00) [] {$\Rightarrow$};
      \end{scope}
      \begin{scope}[xshift=4cm]
        \node (n02) [mycircle] {};
        \node (n01) [mycircle, above of=n02] {};
        \node (n00) [mycircle, above of=n01] {};
        \node (n03) [mycircle, below of=n02] {};
        \node (n04) [mycircle, below of=n03] {};

        \node (n10) [mycircle, left of=n00, node distance=1cm] {};
        \node (n11) [mycircle, below of=n10] {};
        \node (n12) [mycircle, below of=n11] {};
        \node (n13) [mycircle, below of=n12] {};
        \node (n14) [mycircle, below of=n13] {};

        \draw[<-, ultra thick, c1] (n00) -- (n10);
        \draw[<-, ultra thick, c1] (n01) -- (n11);
        \draw[<-, ultra thick, c1] (n02) -- (n12);
        \draw[<-, ultra thick, c1] (n03) -- (n13);
        \draw[<-, ultra thick, c1] (n04) -- (n14);
      \end{scope}
      \begin{scope}[yshift=-3cm]
        \begin{scope}[xshift=0cm]
          \node (n00) [mycircle] {};
          \node (n12) [mycircle, right of=n00, node distance=1cm] {};
          \node (n11) [mycircle, above of=n12] {};
          \node (n10) [mycircle, above of=n11] {};
          \node (n13) [mycircle, below of=n12] {};
          \node (n14) [mycircle, below of=n13] {};

          \draw[->, ultra thick, c1] (n00) -- (n10);
          \draw[->, ultra thick, c1] (n00) -- (n11);
          \draw[->, ultra thick, c1] (n00) -- (n12);
          \draw[->, ultra thick, c1] (n00) -- (n13);
          \draw[->, ultra thick, c1] (n00) -- (n14);
        \end{scope}
        \begin{scope}[xshift=2cm]
          \node (n00) [] {$\Rightarrow$};
        \end{scope}
        \begin{scope}[xshift=4cm]
          \node (n00) [mycircle] {};
          \node (n12) [mycircle, left of=n00, node distance=1cm] {};
          \node (n11) [mycircle, above of=n12] {};
          \node (n10) [mycircle, above of=n11] {};
          \node (n13) [mycircle, below of=n12] {};
          \node (n14) [mycircle, below of=n13] {};

          \draw[<-, ultra thick, c1] (n00) -- (n10);
          \draw[<-, ultra thick, c1] (n00) -- (n11);
          \draw[<-, ultra thick, c1] (n00) -- (n12);
          \draw[<-, ultra thick, c1] (n00) -- (n13);
          \draw[<-, ultra thick, c1] (n00) -- (n14);
        \end{scope}
      \end{scope}
    \end{tikzpicture}
  \end{center}
  \caption{Computational graphs of two \lstinline[style=cstyle,basicstyle=\small\ttfamily]|deep_copy| and the graph of their reverse pass.}
  \label{fig::deep_copy_graph}
\end{figure}
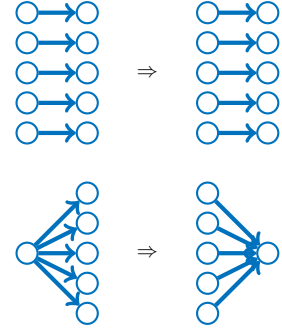

A \codestyle{Kokkos::deep\_copy} can be used to copy the entries of a view to another view or to set all the entries of a View to a given value.
The computational graphs of these two use cases are different as illustrated in Fig.~\ref{fig::deep_copy_graph} and therefore their differentiation needs to be different as well.

As the graph of a \codestyle{deep\_copy} used to copy one view to another is embarrassingly parallel, the reverse pass of the deep copy is also embarrassingly parallel.
In the Kokkos programming model, an embarrassingly parallel region is written using a \codestyle{parallel\_for}.
The reverse pass is therefore written as \codestyle{parallel\_for} as illustrated in Fig.~\ref{fig::deep_copy_views} in the case of 1D views.
In the code of Fig.~\ref{fig::deep_copy_views} and in the rest of this paper \codestyle{KOKKOS_LAMBDA} is a macro used to replace the capture clause of the lambdas.
When a host backend is used, \codestyle{KOKKOS_LAMBDA} is just replaced by a capture-by-value clause \codestyle{[=]} and if the CUDA or HIP backend is used, \codestyle{KOKKOS_LAMBDA} is replaced by \codestyle{[=] __host__ __device__}.

\begin{figure}
  \begin{lstlisting}[style=cstyle]
deep_copy(a_view, b_view);
    \end{lstlisting}
  \begin{center}
    $\Downarrow$
  \end{center}
  \begin{lstlisting}[style=cstyle]
Kokkos::parallel_for(a.extent(0), 
KOKKOS_LAMBDA(const int i) {
    _d_b_view(i) += _d_a_view(i);
});
    \end{lstlisting}
  \caption{Differentiation of \lstinline[style=cstyle,basicstyle=\small\ttfamily]|deep_copy| of one 1D view to another one.}
  \label{fig::deep_copy_views}
\end{figure}

The graph of a \codestyle{deep\_copy} used to set all the entries of a view to a scalar value is a scatter graph.
Therefore, the graph of the reverse pass of the deep copy is a gather graph.
This time, a \codestyle{parallel\_for} cannot be used to efficiently implement the reverse pass as there would be data races between different threads that try to write to the same memory location.
Instead, it is recommended to implement gather graphs using \codestyle{parallel\_reduce} in the Kokkos programming model.
The reverse pass of a \codestyle{deep\_copy} used to set all the entries of a 1D view to a scalar value is illustrated in Fig.~\ref{fig::deep_copy_scalar}.

\begin{figure}
  \begin{lstlisting}[style=cstyle]
deep_copy(a_view, b);
    \end{lstlisting}
  \begin{center}
    $\Downarrow$
  \end{center}
  \begin{lstlisting}[style=cstyle]
Kokkos::parallel_reduce(a.extent(0), 
KOKKOS_LAMBDA (const int i, double &u ) {
    u += _d_a_view(i);
}, _d_b );
    \end{lstlisting}
  \caption{Differentiation of \lstinline[style=cstyle,basicstyle=\small\ttfamily]|deep_copy| of one scalar to one 1D view.}
  \label{fig::deep_copy_scalar}
\end{figure}

Because of the importance of \codestyle{deep\_copy}, we implemented a portable efficient function \codestyle{parallel_sum} of the reverse pass of the \codestyle{deep\_copy} with one specialization for every rank. The source code of the \codestyle{parallel_sum} function is not shown for sake of brevity.
The function \codestyle{parallel_sum} can be used to add all the entries of a view to the corresponding entries of a different view (this correspond to the case of Fig.~\ref{fig::deep_copy_views}) or to sum all the entries of a view and add the result to a scalar value  (this correspond to the case of Fig.~\ref{fig::deep_copy_scalar}).

\subsection{parallel\_for}

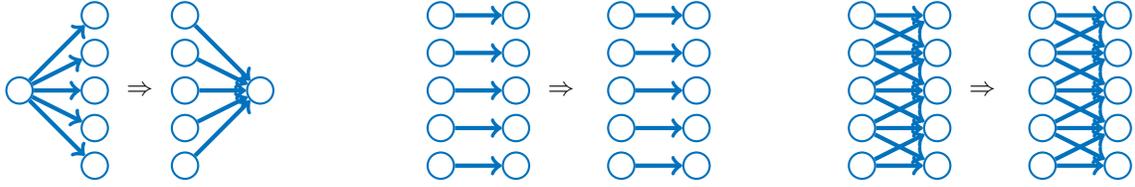
\begin{figure*}
  \begin{center}
    \begin{tikzpicture}[scale=0.8, node distance=0.5cm]
      \begin{scope}[xshift=0cm]
        \node (n00) [mycircle] {};
        \node (n12) [mycircle, right of=n00, node distance=1cm] {};
        \node (n11) [mycircle, above of=n12] {};
        \node (n10) [mycircle, above of=n11] {};
        \node (n13) [mycircle, below of=n12] {};
        \node (n14) [mycircle, below of=n13] {};

        \draw[->, ultra thick, c1] (n00) -- (n10);
        \draw[->, ultra thick, c1] (n00) -- (n11);
        \draw[->, ultra thick, c1] (n00) -- (n12);
        \draw[->, ultra thick, c1] (n00) -- (n13);
        \draw[->, ultra thick, c1] (n00) -- (n14);
      \end{scope}
      \begin{scope}[xshift=2cm]
        \node (n00) [] {$\Rightarrow$};
      \end{scope}
      \begin{scope}[xshift=4cm]
        \node (n00) [mycircle] {};
        \node (n12) [mycircle, left of=n00, node distance=1cm] {};
        \node (n11) [mycircle, above of=n12] {};
        \node (n10) [mycircle, above of=n11] {};
        \node (n13) [mycircle, below of=n12] {};
        \node (n14) [mycircle, below of=n13] {};

        \draw[<-, ultra thick, c1] (n00) -- (n10);
        \draw[<-, ultra thick, c1] (n00) -- (n11);
        \draw[<-, ultra thick, c1] (n00) -- (n12);
        \draw[<-, ultra thick, c1] (n00) -- (n13);
        \draw[<-, ultra thick, c1] (n00) -- (n14);
      \end{scope}
      \begin{scope}[xshift=7cm]
        \node (n02) [mycircle] {};
        \node (n01) [mycircle, above of=n02] {};
        \node (n00) [mycircle, above of=n01] {};
        \node (n03) [mycircle, below of=n02] {};
        \node (n04) [mycircle, below of=n03] {};

        \node (n10) [mycircle, right of=n00, node distance=1cm] {};
        \node (n11) [mycircle, below of=n10] {};
        \node (n12) [mycircle, below of=n11] {};
        \node (n13) [mycircle, below of=n12] {};
        \node (n14) [mycircle, below of=n13] {};

        \draw[->, ultra thick, c1] (n00) -- (n10);
        \draw[->, ultra thick, c1] (n01) -- (n11);
        \draw[->, ultra thick, c1] (n02) -- (n12);
        \draw[->, ultra thick, c1] (n03) -- (n13);
        \draw[->, ultra thick, c1] (n04) -- (n14);
      \end{scope}
      \begin{scope}[xshift=9cm]
        \node (n00) [] {$\Rightarrow$};
      \end{scope}
      \begin{scope}[xshift=10cm]
        \node (n02) [mycircle] {};
        \node (n01) [mycircle, above of=n02] {};
        \node (n00) [mycircle, above of=n01] {};
        \node (n03) [mycircle, below of=n02] {};
        \node (n04) [mycircle, below of=n03] {};

        \node (n10) [mycircle, right of=n00, node distance=1cm] {};
        \node (n11) [mycircle, below of=n10] {};
        \node (n12) [mycircle, below of=n11] {};
        \node (n13) [mycircle, below of=n12] {};
        \node (n14) [mycircle, below of=n13] {};

        \draw[->, ultra thick, c1] (n00) -- (n10);
        \draw[->, ultra thick, c1] (n01) -- (n11);
        \draw[->, ultra thick, c1] (n02) -- (n12);
        \draw[->, ultra thick, c1] (n03) -- (n13);
        \draw[->, ultra thick, c1] (n04) -- (n14);
      \end{scope}
      \begin{scope}[xshift=14cm]
        \node (n02) [mycircle] {};
        \node (n01) [mycircle, above of=n02] {};
        \node (n00) [mycircle, above of=n01] {};
        \node (n03) [mycircle, below of=n02] {};
        \node (n04) [mycircle, below of=n03] {};

        \node (n10) [mycircle, right of=n00, node distance=1cm] {};
        \node (n11) [mycircle, below of=n10] {};
        \node (n12) [mycircle, below of=n11] {};
        \node (n13) [mycircle, below of=n12] {};
        \node (n14) [mycircle, below of=n13] {};

        \draw[->, ultra thick, c1] (n00) -- (n10);
        \draw[->, ultra thick, c1] (n00) -- (n11);
        \draw[->, ultra thick, c1] (n01) -- (n10);
        \draw[->, ultra thick, c1] (n01) -- (n11);
        \draw[->, ultra thick, c1] (n01) -- (n12);
        \draw[->, ultra thick, c1] (n02) -- (n11);
        \draw[->, ultra thick, c1] (n02) -- (n12);
        \draw[->, ultra thick, c1] (n02) -- (n13);
        \draw[->, ultra thick, c1] (n03) -- (n12);
        \draw[->, ultra thick, c1] (n03) -- (n13);
        \draw[->, ultra thick, c1] (n03) -- (n14);
        \draw[->, ultra thick, c1] (n04) -- (n13);
        \draw[->, ultra thick, c1] (n04) -- (n14);
      \end{scope}
      \begin{scope}[xshift=16cm]
        \node (n00) [] {$\Rightarrow$};
      \end{scope}
      \begin{scope}[xshift=17cm]
        \node (n02) [mycircle] {};
        \node (n01) [mycircle, above of=n02] {};
        \node (n00) [mycircle, above of=n01] {};
        \node (n03) [mycircle, below of=n02] {};
        \node (n04) [mycircle, below of=n03] {};

        \node (n10) [mycircle, right of=n00, node distance=1cm] {};
        \node (n11) [mycircle, below of=n10] {};
        \node (n12) [mycircle, below of=n11] {};
        \node (n13) [mycircle, below of=n12] {};
        \node (n14) [mycircle, below of=n13] {};

        \draw[->, ultra thick, c1] (n00) -- (n10);
        \draw[->, ultra thick, c1] (n00) -- (n11);
        \draw[->, ultra thick, c1] (n01) -- (n10);
        \draw[->, ultra thick, c1] (n01) -- (n11);
        \draw[->, ultra thick, c1] (n01) -- (n12);
        \draw[->, ultra thick, c1] (n02) -- (n11);
        \draw[->, ultra thick, c1] (n02) -- (n12);
        \draw[->, ultra thick, c1] (n02) -- (n13);
        \draw[->, ultra thick, c1] (n03) -- (n12);
        \draw[->, ultra thick, c1] (n03) -- (n13);
        \draw[->, ultra thick, c1] (n03) -- (n14);
        \draw[->, ultra thick, c1] (n04) -- (n13);
        \draw[->, ultra thick, c1] (n04) -- (n14);
      \end{scope}
    \end{tikzpicture}
  \end{center}
  \caption{Three examples of computational graphs of Kokkos parallel for and their reverse pass graph.}
  \label{fig::kokkos_parallel_for}
\end{figure*}

The \codestyle{Kokkos::parallel_for} can be used to represent a variety of computational graphs as illustrated in Fig.~\ref{fig::kokkos_parallel_for}.
The reverse pass of a \codestyle{parallel_for} is either another \codestyle{parallel_for} or a \codestyle{parallel_reduce}.
While writing the reverse pass as a \codestyle{Kokkos::parallel_reduce} would improve the performance of the generated code, we think that this situation is unlikely to occur in practice. In order for a reverse pass to be implementable using a \codestyle{parallel_reduce}, the forward graph must only write to views and read from one scalar value.
The reverse pass in this situation is easily implemented using one or more \codestyle{deep\_copy} calls instead.

In some cases a \codestyle{parallel\_for} in the reverse pass causes a race condition where multiple threads might attempt to update a value simultaneously.
We implemented automatic detection for these cases and automatically replace a non-atomic update (\codestyle{+=} operator) with \codestyle{Kokkos::atomic_add}.
The implemented logic flags a view access as potentially not thread-safe if one of the following statement is true:
\begin{itemize}
  \item Inside a parallel region, an index used to access a view is a view access itself for example \codestyle{myView1(myView2(i))},
  \item Inside a parallel region, a view is accessed multiple times with different expressions that include the loop counter variable.
\end{itemize}

\section{Results}
\label{sec::Results}

The most important result of this work is the illustrated feasibility of a compiler-based source code transformation for performance portable gradient computation.

We will now illustrate the performance portability of the compiler-based source code transformation on a Kokkos-based function with multiple kernels illustrated in Fig.~\ref{function_f}.
The function takes two 1D Kokkos views $\boldsymbol{x}$ and $\boldsymbol{b}$ as input and computes a scalar output that correspond to the squared value of the norm of a 1D Laplacian where the vector $\boldsymbol{x}$ is multiplied by $3$:
\[
  f(\boldsymbol{x}, \boldsymbol{b}) = \|\boldsymbol{y}(\boldsymbol{x}, \boldsymbol{b})\|^2,
\]
where $\boldsymbol{y}$ is:
\[
  \boldsymbol{y}(\boldsymbol{x}, \boldsymbol{b}) = \boldsymbol{A}\left(3\,\boldsymbol{x}\right)-\boldsymbol{b},
\]
and where $\boldsymbol{A}$ is:
\[
  \boldsymbol{A} = \begin{bmatrix}
    2  & -1     &        &        &    \\
    -1 & 2      & -1     &        &    \\
       & \ddots & \ddots & \ddots &    \\
       &        & -1     & 2      & -1 \\
       &        &        & -1     & 2
  \end{bmatrix}.
\]

In the code illustrated in Fig.~\ref{function_f}, we use a 2-D view to store \codestyle{y} and \codestyle{y2} that correspond to $\boldsymbol{y}$ and a vector that stores the squared values of $\boldsymbol{y}$, respectively.
We could not compute the power operation in place because our implementation is currently not able to create the tape needed for the reverse mode within a kernel; it can only record values between kernel launches.
The computational graph of the computation of $f$ is illustrated in Fig.~\ref{cg_f}.

\begin{figure*}[ht]
  \begin{lstlisting}[style=cstyle]
template <typename ViewtypeX>
typename ViewtypeX::value_type normRes1DLaplacianSQ(ViewtypeX x, ViewtypeX b) {
  typename ViewtypeX::value_type sum;
  Kokkos::View<typename ViewtypeX::value_type**> y_2D("y_2D", x.extent(0), 2);
  auto y = Kokkos::subview( y_2D, Kokkos::ALL, 0);
  auto y2 = Kokkos::subview( y_2D, Kokkos::ALL, 1);

  Kokkos::parallel_for( x.extent(0), KOKKOS_LAMBDA ( const int j0) {
    x(j0) = 3*x(j0);
  });

  Kokkos::parallel_for( x.extent(0), KOKKOS_LAMBDA ( const int j) {
    y(j) = 2*x(j) - b(j);
    if (j != 0)
      y(j) -= x(j-1);
    if (j != x.extent(0)-1)  
      y(j) -= x(j+1);
    y2(j) = y(j) * y(j);
  });

  kokkos_builtin_derivative::parallel_sum(sum, y2);
  return sum;
}
    \end{lstlisting}
  \caption{Example of a Kokkos-based function consisting of multiple kernels.}
  \label{function_f}
\end{figure*}
\begin{figure}[ht]
  \centering
  \begin{tikzpicture}[scale=0.55, every node/.style={scale=0.55}, node distance=2cm]
    \begin{scope}[xshift=0cm]
      \node (n00) [mycircle] {\codestyle{x(0)}};

      \node (n10)  [mycircle, right of=n00, node distance=2cm] {\codestyle{x(0)}};
      \node (n10b) [mycircle, below of=n10] {\codestyle{b(0)}};
      \node (n11)  [mycircle, below of=n10b] {\codestyle{x(1)}};
      \node (n11b) [mycircle, below of=n11] {\codestyle{b(1)}};
      \node (n12)  [mycircle, below of=n11b] {\codestyle{x(2)}};
      \node (n12b) [mycircle, below of=n12] {\codestyle{b(2)}};

      \node (n01) [mycircle, left of=n11, node distance=2cm] {\codestyle{x(1)}};
      \node (n02) [mycircle, left of=n12, node distance=2cm] {\codestyle{x(2)}};

      \node at ($0.5*(n10)+0.5*(n10b)+(2cm,0)$) (n20) [mycircle] {\codestyle{y2(0)}} ;
      \node at ($0.5*(n11)+0.5*(n11b)+(2cm,0)$) (n21) [mycircle] {\codestyle{y2(1)}};
      \node at ($0.5*(n12)+0.5*(n12b)+(2cm,0)$) (n22) [mycircle] {\codestyle{y2(2)}};

      \node at ($0.5*(n20)+0.5*(n22)+(2cm,0)$) (n30) [mycircle] {\codestyle{sum}};

      \draw[->, ultra thick, c1] (n00) -- (n10);
      \draw[->, ultra thick, c1] (n01) -- (n11);
      \draw[->, ultra thick, c1] (n02) -- (n12);

      \draw[->, ultra thick, c1] (n10b) -- (n20);
      \draw[->, ultra thick, c1] (n11b) -- (n21);
      \draw[->, ultra thick, c1] (n12b) -- (n22);

      \draw[->, ultra thick, c1] (n10) -- (n20);
      \draw[->, ultra thick, c1] (n10) -- (n21);
      \draw[->, ultra thick, c1] (n11) -- (n20);
      \draw[->, ultra thick, c1] (n11) -- (n21);
      \draw[->, ultra thick, c1] (n11) -- (n22);
      \draw[->, ultra thick, c1] (n12) -- (n21);
      \draw[->, ultra thick, c1] (n12) -- (n22);

      \draw[->, ultra thick, c1] (n20) -- (n30);
      \draw[->, ultra thick, c1] (n21) -- (n30);
      \draw[->, ultra thick, c1] (n22) -- (n30);
    \end{scope}
  \end{tikzpicture}
  \caption{Computational graph of the quantity of interest $f$.}
  \label{cg_f}
  \begin{tikzpicture}[scale=0.55, every node/.style={scale=0.55}, node distance=2cm]
    \begin{scope}[xshift=0cm]
      \node (n00) [mycircle] {\codestyle{x(0)}};

      \node (n10)  [mycircle, right of=n00, node distance=2cm] {\codestyle{x(0)}};
      \node (n10b) [mycircle, below of=n10] {\codestyle{b(0)}};
      \node (n11)  [mycircle, below of=n10b] {\codestyle{x(1)}};
      \node (n11b) [mycircle, below of=n11] {\codestyle{b(1)}};
      \node (n12)  [mycircle, below of=n11b] {\codestyle{x(2)}};
      \node (n12b) [mycircle, below of=n12] {\codestyle{b(2)}};

      \node (n01) [mycircle, left of=n11, node distance=2cm] {\codestyle{x(1)}};
      \node (n02) [mycircle, left of=n12, node distance=2cm] {\codestyle{x(2)}};

      \node at ($0.5*(n10)+0.5*(n10b)+(2cm,0)$) (n20) [mycircle] {\codestyle{y2(0)}} ;
      \node at ($0.5*(n11)+0.5*(n11b)+(2cm,0)$) (n21) [mycircle] {\codestyle{y2(1)}};
      \node at ($0.5*(n12)+0.5*(n12b)+(2cm,0)$) (n22) [mycircle] {\codestyle{y2(2)}};

      \node at ($0.5*(n20)+0.5*(n22)+(2cm,0)$) (n30) [mycircle] {\codestyle{sum}};

      \node (n40) [mycircle, right of=n20, node distance=4cm] {\codestyle{_d_y2(0)}};
      \node (n41) [mycircle, right of=n21, node distance=4cm] {\codestyle{_d_y2(1)}};
      \node (n42) [mycircle, right of=n22, node distance=4cm] {\codestyle{_d_y2(2)}};

      \node (n50)  [mycircle, right of=n10,  node distance=8cm] {\codestyle{_d_x(0)}};
      \node (n50b) [mycircle, right of=n10b, node distance=8cm] {\codestyle{_d_b(0)}};
      \node (n51)  [mycircle, right of=n11,  node distance=8cm] {\codestyle{_d_x(1)}};
      \node (n51b) [mycircle, right of=n11b, node distance=8cm] {\codestyle{_d_b(1)}};
      \node (n52)  [mycircle, right of=n12,  node distance=8cm] {\codestyle{_d_x(2)}};
      \node (n52b) [mycircle, right of=n12b, node distance=8cm] {\codestyle{_d_b(2)}};

      \node (n60) [mycircle, right of=n00,  node distance=12cm] {\codestyle{_d_x(0)}};
      \node (n61) [mycircle, right of=n01,  node distance=12cm] {\codestyle{_d_x(1)}};
      \node (n62) [mycircle, right of=n02,  node distance=12cm] {\codestyle{_d_x(2)}};

      \draw[->, ultra thick, c1] (n00) -- (n10);
      \draw[->, ultra thick, c1] (n01) -- (n11);
      \draw[->, ultra thick, c1] (n02) -- (n12);

      \draw[->, ultra thick, c1] (n10b) -- (n20);
      \draw[->, ultra thick, c1] (n11b) -- (n21);
      \draw[->, ultra thick, c1] (n12b) -- (n22);

      \draw[->, ultra thick, c1] (n10) -- (n20);
      \draw[->, ultra thick, c1] (n10) -- (n21);
      \draw[->, ultra thick, c1] (n11) -- (n20);
      \draw[->, ultra thick, c1] (n11) -- (n21);
      \draw[->, ultra thick, c1] (n11) -- (n22);
      \draw[->, ultra thick, c1] (n12) -- (n21);
      \draw[->, ultra thick, c1] (n12) -- (n22);

      \draw[->, ultra thick, c1] (n20) -- (n30);
      \draw[->, ultra thick, c1] (n21) -- (n30);
      \draw[->, ultra thick, c1] (n22) -- (n30);

      \draw[<-, ultra thick, c1] (n60) -- (n50);
      \draw[<-, ultra thick, c1] (n61) -- (n51);
      \draw[<-, ultra thick, c1] (n62) -- (n52);

      \draw[<-, ultra thick, c1] (n50b) -- (n40);
      \draw[<-, ultra thick, c1] (n51b) -- (n41);
      \draw[<-, ultra thick, c1] (n52b) -- (n42);

      \draw[<-, ultra thick, c1] (n50) -- (n40);
      \draw[<-, ultra thick, c1] (n50) -- (n41);
      \draw[<-, ultra thick, c1] (n51) -- (n40);
      \draw[<-, ultra thick, c1] (n51) -- (n41);
      \draw[<-, ultra thick, c1] (n51) -- (n42);
      \draw[<-, ultra thick, c1] (n52) -- (n41);
      \draw[<-, ultra thick, c1] (n52) -- (n42);

      \draw[<-, ultra thick, c1] (n40) -- (n30);
      \draw[<-, ultra thick, c1] (n41) -- (n30);
      \draw[<-, ultra thick, c1] (n42) -- (n30);
    \end{scope}
  \end{tikzpicture}
  \caption{Computational graph of the gradient of the quantity of interest $f$ including both the forward and reverse passes.}
  \label{cg_gf}
\end{figure}

We are interested in computing the gradient of the $f$ with respect to $\boldsymbol{x}$. The function contains a sequence of three Kokkos kernels:
\begin{enumerate}
  \item A linear kernel that multiplies in place all the entries of $\boldsymbol{x}$ by $3$.
  \item A non-linear kernel that computes $\boldsymbol{y} = \boldsymbol{A}\boldsymbol{x}-\boldsymbol{b}$ given $\boldsymbol{x}$ and $\boldsymbol{b}$ and that squares element-wise all the entries of $\boldsymbol{y}$.
  \item A parallel reduction over all the entries of the vector $\boldsymbol{y}$.
\end{enumerate}

The generated code using the modified Kokkos-aware Clad for the gradient computation of the function of Fig.~\ref{function_f} is listed in Fig.~\ref{grad_f}.
The generation has been tested on an Apple M2 ARM CPU and on Intel CPUs.

\begin{figure*}[htbp]
  \begin{lstlisting}[style=cstyle]
template <typename type_x> 
void normRes1DLaplacianSQ_grad(type_x x, type_x b, type_x _d_x, type_x _d_b) {
    typename type_x::value_type _d_sum = 0;
    Kokkos::View<typename type_x::value_type **> _d_y_2D("_d_y_2D", x.extent(0), 2);
    auto _d_y = Kokkos::subview(_d_y_2D, Kokkos::ALL, 0);
    auto _d_y2 = Kokkos::subview(_d_y_2D, Kokkos::ALL, 1);
    typename type_x::value_type sum;
    Kokkos::View<typename type_x::value_type **> y_2D("y_2D", x.extent(0), 2);
    auto y = Kokkos::subview(y_2D, Kokkos::ALL, 0);
    auto y2 = Kokkos::subview(y_2D, Kokkos::ALL, 1);
    Kokkos::parallel_for(x.extent(0), KOKKOS_LAMBDA(const int j0) {
        x(j0) = 3 * x(j0);
    });
    Kokkos::parallel_for(x.extent(0), KOKKOS_LAMBDA(const int j) {
        y(j) = 2 * x(j) - b(j);
        if (j != 0)
            y(j) -= x(j - 1);
        if (j != x.extent(0) - 1)
            y(j) -= x(j + 1);
        y2(j) = y(j) * y(j);
    });
    kokkos_builtin_derivative::parallel_sum(sum, y2);
    goto _label0;
  _label0:
    _d_sum += 1;
    kokkos_builtin_derivative::parallel_sum(_d_y2, _d_sum);
    {
        Kokkos::parallel_for(x.extent(0), KOKKOS_LAMBDA(const int j) {
            {
                double _r_d4 = _d_y2(j);
                _d_y2(j) -= _r_d4;
                _d_y(j) += _r_d4 * y(j);
                _d_y(j) += y(j) * _r_d4;
                _d_y2(j);
            }
            if (j != x.extent(0) - 1) {
                double _r_d3 = _d_y(j);
                Kokkos::atomic_add(&_d_x(j + 1), -_r_d3);
                _d_y(j);
            }
            if (j != 0) {
                double _r_d2 = _d_y(j);
                Kokkos::atomic_add(&_d_x(j - 1), -_r_d2);
                _d_y(j);
            }
            {
                double _r_d1 = _d_y(j);
                _d_y(j) -= _r_d1;
                Kokkos::atomic_add(&_d_x(j), 2 * _r_d1);
                _d_b(j) += -_r_d1;
                _d_y(j);
            }
        });
    }
    Kokkos::parallel_for(x.extent(0), KOKKOS_LAMBDA(const int j0) {
        {
            double _r_d0 = _d_x(j0);
            _d_x(j0) -= _r_d0;
            _d_x(j0) += 3 * _r_d0;
            _d_x(j0);
        }
    });
}
    \end{lstlisting}
  \caption{Automatically generated gradient of the function of Fig.~\ref{function_f}.}
  \label{grad_f}
\end{figure*}

There are several aspects that need to be highlighted in the generated function:
\begin{enumerate}
  \item There are two more input arguments which are Kokkos views used to pass the gradient results to the caller of the function.
  \item The code has three main parts: the initialization phase where Kokkos views and other temporary variables are initialized,
        the forward pass that reproduces the same operations in the same order as Fig.~\ref{function_f}, and the reverse pass.
  \item For each kernel of the function of Fig.~\ref{function_f} there are two kernels in Fig.~\ref{grad_f}: one in the forward pass and one in the reverse pass.
  \item If the same view entry is read by multiple threads within the same kernel of the input source code, the reverse pass will have multiple threads that want to write to a same view entry.
        Those data races are identified by the Kokkos-aware Clad and atomic operations are introduced in the reverse pass.
\end{enumerate}

\begin{figure}[p]
  \centering
  \pgfplotsset{scaled x ticks=false}
  \begin{tikzpicture}

    \begin{axis}[
        xlabel={$\mathrm{dim}(\boldsymbol{x})$},
        ylabel={ Time [sec]},
        xmin=-1e2, xmax=1e4,
        ymin=0, ymax=1e-4,
        xtick={1, 2e3, 4e3, 6e3, 8e3, 1e4},
        compat=1.3,
        width=8cm,
        height=5.5cm,
        tick align=outside,
        tick pos=left,
        xmajorgrids,
        x grid style={white},
        ymajorgrids,
        grid=both,
        y grid style={white},
        clip marker paths=true,
        axis line style={white},
        axis background/.style={fill=white!89.803921568627459!black},
        name=boundary,
        legend style={nodes={scale=0.75, transform shape}, fill opacity=0.8, draw opacity=1, text opacity=1, draw=white!90!black, at={(.99,.01)}, anchor=south east},
      ]
      \addplot [thick, c1]
      table [col sep=space] {
          1.000000000000000000e+00 4.178600000000000277e-05
          1.000000000000000000e+02 4.252549999999999798e-05
          2.000000000000000000e+02 4.289599999999999723e-05
          3.000000000000000000e+02 4.360850000000000099e-05
          4.000000000000000000e+02 4.397799999999999805e-05
          5.000000000000000000e+02 4.369999999999999834e-05
          6.000000000000000000e+02 4.422100000000000233e-05
          7.000000000000000000e+02 4.374850000000000304e-05
          8.000000000000000000e+02 4.426299999999999956e-05
          9.000000000000000000e+02 4.387549999999999691e-05
          1.000000000000000000e+03 4.442500000000000241e-05
          1.100000000000000000e+03 4.426650000000000045e-05
          1.200000000000000000e+03 4.364799999999999952e-05
          1.300000000000000000e+03 4.425500000000000234e-05
          1.400000000000000000e+03 4.464050000000000060e-05
          1.500000000000000000e+03 4.435050000000000168e-05
          1.600000000000000000e+03 4.439849999999999851e-05
          1.700000000000000000e+03 4.432699999999999759e-05
          1.800000000000000000e+03 4.466749999999999882e-05
          1.900000000000000000e+03 4.454299999999999688e-05
          2.000000000000000000e+03 4.459450000000000138e-05
          2.100000000000000000e+03 4.420949999999999744e-05
          2.200000000000000000e+03 4.463349999999999880e-05
          2.300000000000000000e+03 4.426650000000000045e-05
          2.400000000000000000e+03 4.451150000000000234e-05
          2.500000000000000000e+03 4.763149999999999958e-05
          2.600000000000000000e+03 4.791149999999999690e-05
          2.700000000000000000e+03 4.788199999999999997e-05
          2.800000000000000000e+03 4.794500000000000259e-05
          2.900000000000000000e+03 4.784099999999999816e-05
          3.000000000000000000e+03 4.788650000000000307e-05
          3.100000000000000000e+03 4.778650000000000063e-05
          3.200000000000000000e+03 4.781850000000000304e-05
          3.300000000000000000e+03 4.788699999999999739e-05
          3.400000000000000000e+03 4.789900000000000337e-05
          3.500000000000000000e+03 4.785450000000000066e-05
          3.600000000000000000e+03 4.774000000000000032e-05
          3.700000000000000000e+03 4.784000000000000275e-05
          3.800000000000000000e+03 4.760400000000000026e-05
          3.900000000000000000e+03 4.793500000000000099e-05
          4.000000000000000000e+03 4.767749999999999880e-05
          4.100000000000000000e+03 4.795700000000000180e-05
          4.200000000000000000e+03 4.785200000000000196e-05
          4.300000000000000000e+03 4.799249999999999832e-05
          4.400000000000000000e+03 4.785999999999999917e-05
          4.500000000000000000e+03 4.786150000000000246e-05
          4.600000000000000000e+03 4.784900000000000215e-05
          4.700000000000000000e+03 4.784000000000000275e-05
          4.800000000000000000e+03 4.787650000000000147e-05
          4.900000000000000000e+03 4.790150000000000207e-05
          5.000000000000000000e+03 4.760299999999999807e-05
          5.100000000000000000e+03 4.784399999999999797e-05
          5.200000000000000000e+03 4.786649999999999987e-05
          5.300000000000000000e+03 4.780699999999999815e-05
          5.400000000000000000e+03 4.767650000000000338e-05
          5.500000000000000000e+03 4.764550000000000317e-05
          5.600000000000000000e+03 4.786899999999999857e-05
          5.700000000000000000e+03 4.795000000000000000e-05
          5.800000000000000000e+03 4.777349999999999923e-05
          5.900000000000000000e+03 4.798500000000000221e-05
          6.000000000000000000e+03 4.778750000000000283e-05
          6.100000000000000000e+03 4.770400000000000269e-05
          6.200000000000000000e+03 4.783300000000000095e-05
          6.300000000000000000e+03 4.792999999999999680e-05
          6.400000000000000000e+03 4.780149999999999964e-05
          6.500000000000000000e+03 4.778050000000000103e-05
          6.600000000000000000e+03 4.778249999999999864e-05
          6.700000000000000000e+03 4.783900000000000056e-05
          6.800000000000000000e+03 4.778199999999999754e-05
          6.900000000000000000e+03 4.787600000000000037e-05
          7.000000000000000000e+03 4.787700000000000256e-05
          7.100000000000000000e+03 4.784999999999999757e-05
          7.200000000000000000e+03 4.774799999999999753e-05
          7.300000000000000000e+03 4.791650000000000108e-05
          7.400000000000000000e+03 4.793999999999999840e-05
          7.500000000000000000e+03 4.780149999999999964e-05
          7.600000000000000000e+03 4.782400000000000155e-05
          7.700000000000000000e+03 4.775499999999999933e-05
          7.800000000000000000e+03 4.813199999999999927e-05
          7.900000000000000000e+03 4.794949999999999890e-05
          8.000000000000000000e+03 4.777949999999999884e-05
          8.100000000000000000e+03 4.768699999999999930e-05
          8.200000000000000000e+03 4.787350000000000166e-05
          8.300000000000000000e+03 4.763500000000000048e-05
          8.400000000000000000e+03 4.780699999999999815e-05
          8.500000000000000000e+03 4.775250000000000062e-05
          8.600000000000000000e+03 4.784600000000000235e-05
          8.700000000000000000e+03 4.786700000000000096e-05
          8.800000000000000000e+03 4.772150000000000041e-05
          8.900000000000000000e+03 4.785750000000000046e-05
          9.000000000000000000e+03 4.782799999999999676e-05
          9.100000000000000000e+03 4.787650000000000147e-05
          9.200000000000000000e+03 4.784099999999999816e-05
          9.300000000000000000e+03 4.773649999999999942e-05
          9.400000000000000000e+03 4.770799999999999791e-05
          9.500000000000000000e+03 4.783100000000000334e-05
          9.600000000000000000e+03 4.789900000000000337e-05
          9.700000000000000000e+03 4.781049999999999905e-05
          9.800000000000000000e+03 4.778750000000000283e-05
          9.900000000000000000e+03 4.776099999999999893e-05
          1.000000000000000000e+04 4.782199999999999716e-05
        };
      \addlegendentry{$f$}

      \addplot [thick, c2]
      table [col sep=space] {
          1.000000000000000000e+00 7.470049999999999401e-05
          1.000000000000000000e+02 7.622300000000000527e-05
          2.000000000000000000e+02 7.681000000000000490e-05
          3.000000000000000000e+02 7.806949999999999513e-05
          4.000000000000000000e+02 7.827500000000000527e-05
          5.000000000000000000e+02 7.813900000000000522e-05
          6.000000000000000000e+02 7.867000000000000403e-05
          7.000000000000000000e+02 7.890400000000000213e-05
          8.000000000000000000e+02 7.915999999999999426e-05
          9.000000000000000000e+02 7.900150000000000586e-05
          1.000000000000000000e+03 7.948850000000000306e-05
          1.100000000000000000e+03 7.934099999999999812e-05
          1.200000000000000000e+03 7.885000000000000570e-05
          1.300000000000000000e+03 7.947150000000000644e-05
          1.400000000000000000e+03 7.912049999999999574e-05
          1.500000000000000000e+03 7.942050000000000303e-05
          1.600000000000000000e+03 7.960349999999999772e-05
          1.700000000000000000e+03 7.967899999999999386e-05
          1.800000000000000000e+03 7.962149999999999653e-05
          1.900000000000000000e+03 7.973849999999999558e-05
          2.000000000000000000e+03 7.980549999999999341e-05
          2.100000000000000000e+03 7.964700000000000501e-05
          2.200000000000000000e+03 7.930299999999999611e-05
          2.300000000000000000e+03 7.962350000000000092e-05
          2.400000000000000000e+03 7.978200000000000287e-05
          2.500000000000000000e+03 8.496999999999999453e-05
          2.600000000000000000e+03 8.501800000000000492e-05
          2.700000000000000000e+03 8.548100000000000370e-05
          2.800000000000000000e+03 8.529299999999999805e-05
          2.900000000000000000e+03 8.543999999999999512e-05
          3.000000000000000000e+03 8.557650000000000304e-05
          3.100000000000000000e+03 8.536449999999999897e-05
          3.200000000000000000e+03 8.534449999999999578e-05
          3.300000000000000000e+03 8.514450000000000447e-05
          3.400000000000000000e+03 8.538949999999999958e-05
          3.500000000000000000e+03 8.538499999999999649e-05
          3.600000000000000000e+03 8.533849999999999617e-05
          3.700000000000000000e+03 8.523550000000000072e-05
          3.800000000000000000e+03 8.534099999999999488e-05
          3.900000000000000000e+03 8.509100000000000236e-05
          4.000000000000000000e+03 8.507200000000000135e-05
          4.100000000000000000e+03 8.525349999999999953e-05
          4.200000000000000000e+03 8.550999999999999953e-05
          4.300000000000000000e+03 8.543299999999999332e-05
          4.400000000000000000e+03 8.510699999999999678e-05
          4.500000000000000000e+03 8.524600000000000341e-05
          4.600000000000000000e+03 8.542449999999999501e-05
          4.700000000000000000e+03 8.554949999999999805e-05
          4.800000000000000000e+03 8.536300000000000246e-05
          4.900000000000000000e+03 8.550099999999999335e-05
          5.000000000000000000e+03 8.530999999999999467e-05
          5.100000000000000000e+03 8.505750000000000344e-05
          5.200000000000000000e+03 8.509149999999999668e-05
          5.300000000000000000e+03 8.525200000000000302e-05
          5.400000000000000000e+03 8.519500000000000000e-05
          5.500000000000000000e+03 8.534050000000000056e-05
          5.600000000000000000e+03 8.525499999999999604e-05
          5.700000000000000000e+03 8.550350000000000561e-05
          5.800000000000000000e+03 8.558449999999999348e-05
          5.900000000000000000e+03 8.528400000000000542e-05
          6.000000000000000000e+03 8.506250000000000085e-05
          6.100000000000000000e+03 8.532799999999999348e-05
          6.200000000000000000e+03 8.523099999999999763e-05
          6.300000000000000000e+03 8.556800000000000473e-05
          6.400000000000000000e+03 8.540049999999999660e-05
          6.500000000000000000e+03 8.535299999999999409e-05
          6.600000000000000000e+03 8.510500000000000595e-05
          6.700000000000000000e+03 8.548749999999999763e-05
          6.800000000000000000e+03 8.534349999999999358e-05
          6.900000000000000000e+03 8.548049999999999583e-05
          7.000000000000000000e+03 8.523599999999999504e-05
          7.100000000000000000e+03 8.550000000000000471e-05
          7.200000000000000000e+03 8.533949999999999837e-05
          7.300000000000000000e+03 8.544400000000000389e-05
          7.400000000000000000e+03 8.555699999999999417e-05
          7.500000000000000000e+03 8.550750000000000082e-05
          7.600000000000000000e+03 8.534549999999999797e-05
          7.700000000000000000e+03 8.549699999999999813e-05
          7.800000000000000000e+03 8.556049999999999506e-05
          7.900000000000000000e+03 8.546949999999999882e-05
          8.000000000000000000e+03 8.534400000000000146e-05
          8.100000000000000000e+03 8.547949999999999364e-05
          8.200000000000000000e+03 8.563049999999999948e-05
          8.300000000000000000e+03 8.550199999999999554e-05
          8.400000000000000000e+03 8.524100000000000600e-05
          8.500000000000000000e+03 8.557599999999999517e-05
          8.600000000000000000e+03 8.548950000000000201e-05
          8.700000000000000000e+03 8.554550000000000283e-05
          8.800000000000000000e+03 8.537799999999999469e-05
          8.900000000000000000e+03 8.545649999999999742e-05
          9.000000000000000000e+03 8.560349999999999448e-05
          9.100000000000000000e+03 8.558899999999999657e-05
          9.200000000000000000e+03 8.542699999999999372e-05
          9.300000000000000000e+03 8.542950000000000598e-05
          9.400000000000000000e+03 8.543849999999999860e-05
          9.500000000000000000e+03 8.552949999999999485e-05
          9.600000000000000000e+03 8.545500000000000090e-05
          9.700000000000000000e+03 8.570800000000000001e-05
          9.800000000000000000e+03 8.542850000000000378e-05
          9.900000000000000000e+03 8.543149999999999681e-05
          1.000000000000000000e+04 8.544349999999999602e-05

        };
      \addlegendentry{$\nabla f$ with source code transformation}

    \end{axis}
  \end{tikzpicture}
  \vspace*{-1.05cm}
  \caption{Wall clock time on H100 of evaluating the function of Fig.~\ref{function_f} and the generated gradient using source code transformation.}
  \label{perf_results_H100}

  \pgfplotsset{scaled x ticks=false}
  \begin{tikzpicture}

    \begin{axis}[
        xlabel={$\mathrm{dim}(\boldsymbol{x})$},
        ylabel={ Time [sec]},
        xmin=-1e2, xmax=1e4,
        ymin=0, ymax=2.5e-4,
        xtick={1, 2e3, 4e3, 6e3, 8e3, 1e4},
        compat=1.3,
        width=8cm,
        height=5.5cm,
        tick align=outside,
        tick pos=left,
        xmajorgrids,
        x grid style={white},
        ymajorgrids,
        grid=both,
        y grid style={white},
        clip marker paths=true,
        axis line style={white},
        axis background/.style={fill=white!89.803921568627459!black},
        name=boundary,
        legend style={nodes={scale=0.75, transform shape}, fill opacity=0.8, draw opacity=1, text opacity=1, draw=white!90!black, at={(.99,.01)}, anchor=south east},
      ]

      \addplot [thick, c1]
      table [col sep=space] {1.000000000000000000e+00 8.966899999999999816e-05
          1.000000000000000000e+02 9.859049999999999737e-05
          2.000000000000000000e+02 1.002139999999999997e-04
          3.000000000000000000e+02 1.016310000000000032e-04
          4.000000000000000000e+02 1.049930000000000026e-04
          5.000000000000000000e+02 1.045420000000000030e-04
          6.000000000000000000e+02 1.051379999999999952e-04
          7.000000000000000000e+02 1.052030000000000022e-04
          8.000000000000000000e+02 1.063059999999999949e-04
          9.000000000000000000e+02 1.064400000000000041e-04
          1.000000000000000000e+03 1.074269999999999999e-04
          1.100000000000000000e+03 1.073530000000000003e-04
          1.200000000000000000e+03 1.070169999999999954e-04
          1.300000000000000000e+03 1.070820000000000024e-04
          1.400000000000000000e+03 1.079639999999999983e-04
          1.500000000000000000e+03 1.075280000000000045e-04
          1.600000000000000000e+03 1.076279999999999934e-04
          1.700000000000000000e+03 1.076080000000000038e-04
          1.800000000000000000e+03 1.077829999999999945e-04
          1.900000000000000000e+03 1.074830000000000007e-04
          2.000000000000000000e+03 1.084139999999999957e-04
          2.100000000000000000e+03 1.083649999999999967e-04
          2.200000000000000000e+03 1.075729999999999948e-04
          2.300000000000000000e+03 1.078579999999999963e-04
          2.400000000000000000e+03 1.082140000000000044e-04
          2.500000000000000000e+03 1.083089999999999959e-04
          2.600000000000000000e+03 1.091259999999999984e-04
          2.700000000000000000e+03 1.092760000000000020e-04
          2.800000000000000000e+03 1.089250000000000049e-04
          2.900000000000000000e+03 1.101929999999999934e-04
          3.000000000000000000e+03 1.100930000000000046e-04
          3.100000000000000000e+03 1.094269999999999943e-04
          3.200000000000000000e+03 1.088760000000000059e-04
          3.300000000000000000e+03 1.096369999999999940e-04
          3.400000000000000000e+03 1.095459999999999978e-04
          3.500000000000000000e+03 1.098229999999999953e-04
          3.600000000000000000e+03 1.100320000000000063e-04
          3.700000000000000000e+03 1.103429999999999971e-04
          3.800000000000000000e+03 1.105730000000000000e-04
          3.900000000000000000e+03 1.111089999999999962e-04
          4.000000000000000000e+03 1.113350000000000038e-04
          4.100000000000000000e+03 1.110789999999999982e-04
          4.200000000000000000e+03 1.103729999999999951e-04
          4.300000000000000000e+03 1.106379999999999934e-04
          4.400000000000000000e+03 1.103580000000000029e-04
          4.500000000000000000e+03 1.104529999999999943e-04
          4.600000000000000000e+03 1.105229999999999987e-04
          4.700000000000000000e+03 1.111850000000000002e-04
          4.800000000000000000e+03 1.120309999999999985e-04
          4.900000000000000000e+03 1.125619999999999973e-04
          5.000000000000000000e+03 1.128180000000000030e-04
          5.100000000000000000e+03 1.122920000000000016e-04
          5.200000000000000000e+03 1.121570000000000038e-04
          5.300000000000000000e+03 1.123709999999999987e-04
          5.400000000000000000e+03 1.122469999999999978e-04
          5.500000000000000000e+03 1.121760000000000048e-04
          5.600000000000000000e+03 1.121570000000000038e-04
          5.700000000000000000e+03 1.123269999999999971e-04
          5.800000000000000000e+03 1.132589999999999942e-04
          5.900000000000000000e+03 1.141110000000000057e-04
          6.000000000000000000e+03 1.126719999999999946e-04
          6.100000000000000000e+03 1.129829999999999989e-04
          6.200000000000000000e+03 1.136450000000000004e-04
          6.300000000000000000e+03 1.138049999999999988e-04
          6.400000000000000000e+03 1.134990000000000055e-04
          6.500000000000000000e+03 1.145459999999999973e-04
          6.600000000000000000e+03 1.144310000000000027e-04
          6.700000000000000000e+03 1.145660000000000005e-04
          6.800000000000000000e+03 1.160489999999999997e-04
          6.900000000000000000e+03 1.158189999999999968e-04
          7.000000000000000000e+03 1.156380000000000065e-04
          7.100000000000000000e+03 1.159489999999999973e-04
          7.200000000000000000e+03 1.149570000000000041e-04
          7.300000000000000000e+03 1.156180000000000033e-04
          7.400000000000000000e+03 1.129629999999999957e-04
          7.500000000000000000e+03 1.123759999999999961e-04
          7.600000000000000000e+03 1.126770000000000056e-04
          7.700000000000000000e+03 1.163449999999999982e-04
          7.800000000000000000e+03 1.161539999999999995e-04
          7.900000000000000000e+03 1.143159999999999944e-04
          8.000000000000000000e+03 1.150870000000000045e-04
          8.100000000000000000e+03 1.154979999999999977e-04
          8.200000000000000000e+03 1.164489999999999959e-04
          8.300000000000000000e+03 1.108689999999999985e-04
          8.400000000000000000e+03 1.108739999999999959e-04
          8.500000000000000000e+03 1.064760000000000017e-04
          8.600000000000000000e+03 1.095920000000000038e-04
          8.700000000000000000e+03 1.116500000000000034e-04
          8.800000000000000000e+03 1.128079999999999946e-04
          8.900000000000000000e+03 1.093359999999999981e-04
          9.000000000000000000e+03 1.134689999999999939e-04
          9.100000000000000000e+03 1.134740000000000049e-04
          9.200000000000000000e+03 1.124220000000000021e-04
          9.300000000000000000e+03 1.137549999999999976e-04
          9.400000000000000000e+03 1.164890000000000023e-04
          9.500000000000000000e+03 1.128569999999999937e-04
          9.600000000000000000e+03 1.128680000000000042e-04
          9.700000000000000000e+03 1.132190000000000014e-04
          9.800000000000000000e+03 1.133289999999999986e-04
          9.900000000000000000e+03 1.132880000000000036e-04
          1.000000000000000000e+04 1.136239999999999950e-04
        };
      \addlegendentry{$f$}

      \addplot [thick, c2]
      table [col sep=space] {1.000000000000000000e+00 1.946770000000000065e-04
          1.000000000000000000e+02 1.958650000000000094e-04
          2.000000000000000000e+02 1.997470000000000105e-04
          3.000000000000000000e+02 2.058779999999999963e-04
          4.000000000000000000e+02 2.088330000000000112e-04
          5.000000000000000000e+02 2.102770000000000062e-04
          6.000000000000000000e+02 2.095750000000000119e-04
          7.000000000000000000e+02 2.096359999999999966e-04
          8.000000000000000000e+02 2.118440000000000134e-04
          9.000000000000000000e+02 2.139130000000000100e-04
          1.000000000000000000e+03 2.170239999999999875e-04
          1.100000000000000000e+03 2.151250000000000113e-04
          1.200000000000000000e+03 2.151199999999999868e-04
          1.300000000000000000e+03 2.156419999999999930e-04
          1.400000000000000000e+03 2.161470000000000025e-04
          1.500000000000000000e+03 2.150909999999999910e-04
          1.600000000000000000e+03 2.163479999999999960e-04
          1.700000000000000000e+03 2.149409999999999873e-04
          1.800000000000000000e+03 2.144489999999999927e-04
          1.900000000000000000e+03 2.167290000000000048e-04
          2.000000000000000000e+03 2.161979999999999924e-04
          2.100000000000000000e+03 2.150900000000000023e-04
          2.200000000000000000e+03 2.180609999999999981e-04
          2.300000000000000000e+03 2.160519999999999975e-04
          2.400000000000000000e+03 2.166389999999999972e-04
          2.500000000000000000e+03 2.172999999999999964e-04
          2.600000000000000000e+03 2.182859999999999900e-04
          2.700000000000000000e+03 2.192030000000000085e-04
          2.800000000000000000e+03 2.186470000000000091e-04
          2.900000000000000000e+03 2.201249999999999973e-04
          3.000000000000000000e+03 2.213069999999999870e-04
          3.100000000000000000e+03 2.199849999999999885e-04
          3.200000000000000000e+03 2.199799999999999911e-04
          3.300000000000000000e+03 2.210620000000000055e-04
          3.400000000000000000e+03 2.189030000000000012e-04
          3.500000000000000000e+03 2.206959999999999890e-04
          3.600000000000000000e+03 2.216030000000000126e-04
          3.700000000000000000e+03 2.210419999999999887e-04
          3.800000000000000000e+03 2.211469999999999885e-04
          3.900000000000000000e+03 2.229199999999999867e-04
          4.000000000000000000e+03 2.235519999999999901e-04
          4.100000000000000000e+03 2.241730000000000101e-04
          4.200000000000000000e+03 2.234770000000000018e-04
          4.300000000000000000e+03 2.234969999999999915e-04
          4.400000000000000000e+03 2.238270000000000103e-04
          4.500000000000000000e+03 2.244790000000000034e-04
          4.600000000000000000e+03 2.237960000000000101e-04
          4.700000000000000000e+03 2.237910000000000127e-04
          4.800000000000000000e+03 2.243679999999999904e-04
          4.900000000000000000e+03 2.263819999999999884e-04
          5.000000000000000000e+03 2.260309999999999913e-04
          5.100000000000000000e+03 2.250739999999999935e-04
          5.200000000000000000e+03 2.262459999999999883e-04
          5.300000000000000000e+03 2.263560000000000127e-04
          5.400000000000000000e+03 2.251440000000000114e-04
          5.500000000000000000e+03 2.275740000000000000e-04
          5.600000000000000000e+03 2.267079999999999985e-04
          5.700000000000000000e+03 2.279899999999999906e-04
          5.800000000000000000e+03 2.284759999999999992e-04
          5.900000000000000000e+03 2.290570000000000128e-04
          6.000000000000000000e+03 2.294230000000000021e-04
          6.100000000000000000e+03 2.289620000000000077e-04
          6.200000000000000000e+03 2.292670000000000124e-04
          6.300000000000000000e+03 2.291869999999999997e-04
          6.400000000000000000e+03 2.302639999999999895e-04
          6.500000000000000000e+03 2.309860000000000006e-04
          6.600000000000000000e+03 2.307839999999999913e-04
          6.700000000000000000e+03 2.316970000000000010e-04
          6.800000000000000000e+03 2.329189999999999971e-04
          6.900000000000000000e+03 2.328800000000000065e-04
          7.000000000000000000e+03 2.341219999999999922e-04
          7.100000000000000000e+03 2.333000000000000059e-04
          7.200000000000000000e+03 2.330490000000000111e-04
          7.300000000000000000e+03 2.271080000000000082e-04
          7.400000000000000000e+03 2.263319999999999872e-04
          7.500000000000000000e+03 2.272440000000000083e-04
          7.600000000000000000e+03 2.320719999999999966e-04
          7.700000000000000000e+03 2.358550000000000110e-04
          7.800000000000000000e+03 2.364009999999999885e-04
          7.900000000000000000e+03 2.357690000000000122e-04
          8.000000000000000000e+03 2.362610000000000068e-04
          8.100000000000000000e+03 2.366059999999999908e-04
          8.200000000000000000e+03 2.271130000000000056e-04
          8.300000000000000000e+03 2.285160000000000056e-04
          8.400000000000000000e+03 2.278950000000000127e-04
          8.500000000000000000e+03 2.180609999999999981e-04
          8.600000000000000000e+03 2.275439999999999885e-04
          8.700000000000000000e+03 2.309550000000000004e-04
          8.800000000000000000e+03 2.241929999999999997e-04
          8.900000000000000000e+03 2.320079999999999918e-04
          9.000000000000000000e+03 2.319079999999999894e-04
          9.100000000000000000e+03 2.317119999999999933e-04
          9.200000000000000000e+03 2.327189999999999923e-04
          9.300000000000000000e+03 2.335709999999999902e-04
          9.400000000000000000e+03 2.347279999999999929e-04
          9.500000000000000000e+03 2.320880000000000046e-04
          9.600000000000000000e+03 2.324530000000000053e-04
          9.700000000000000000e+03 2.323229999999999913e-04
          9.800000000000000000e+03 2.333199999999999955e-04
          9.900000000000000000e+03 2.332599999999999995e-04
          1.000000000000000000e+04 2.338209999999999963e-04
        };
      \addlegendentry{$\nabla f$ with source code transformation}

    \end{axis}
  \end{tikzpicture}
  \vspace*{-1.05cm}
  \caption{Wall clock time on MI250X of evaluating the function of Fig.~\ref{function_f} and the generated gradient using source code transformation.}
  \label{perf_results_mi250x}

  \pgfplotsset{scaled x ticks=false}
  \begin{tikzpicture}

    \begin{axis}[
        xlabel={$\mathrm{dim}(\boldsymbol{x})$},
        ylabel={ Time [sec]},
        xmin=-1e2, xmax=1e4,
        ymin=0, ymax=7e-5,
        xtick={1, 2e3, 4e3, 6e3, 8e3, 1e4},
        compat=1.3,
        width=8cm,
        height=5.5cm,
        tick align=outside,
        tick pos=left,
        xmajorgrids,
        x grid style={white},
        ymajorgrids,
        grid=both,
        y grid style={white},
        clip marker paths=true,
        axis line style={white},
        axis background/.style={fill=white!89.803921568627459!black},
        name=boundary,
        legend style={nodes={scale=0.75, transform shape}, fill opacity=0.8, draw opacity=1, text opacity=1, draw=white!90!black, at={(.99,.01)}, anchor=south east},
      ]

      \addplot [thick, c1]
      table [col sep=space] {1.000000000000000000e+00 2.816000000000000144e-05
          1.000000000000000000e+02 3.327649999999999895e-05
          2.000000000000000000e+02 3.400400000000000173e-05
          3.000000000000000000e+02 3.564800000000000157e-05
          4.000000000000000000e+02 3.597349999999999702e-05
          5.000000000000000000e+02 3.576000000000000321e-05
          6.000000000000000000e+02 3.863849999999999945e-05
          7.000000000000000000e+02 3.852850000000000220e-05
          8.000000000000000000e+02 3.924099999999999919e-05
          9.000000000000000000e+02 3.921800000000000297e-05
          1.000000000000000000e+03 3.911850000000000163e-05
          1.100000000000000000e+03 3.868300000000000216e-05
          1.200000000000000000e+03 3.847749999999999879e-05
          1.300000000000000000e+03 3.871750000000000327e-05
          1.400000000000000000e+03 3.844949999999999838e-05
          1.500000000000000000e+03 3.857899999999999773e-05
          1.600000000000000000e+03 3.857100000000000052e-05
          1.700000000000000000e+03 3.836649999999999935e-05
          1.800000000000000000e+03 3.872449999999999829e-05
          1.900000000000000000e+03 3.837799999999999746e-05
          2.000000000000000000e+03 3.844299999999999768e-05
          2.100000000000000000e+03 3.808300000000000113e-05
          2.200000000000000000e+03 3.843149999999999957e-05
          2.300000000000000000e+03 3.817699999999999718e-05
          2.400000000000000000e+03 3.863199999999999875e-05
          2.500000000000000000e+03 3.860099999999999854e-05
          2.600000000000000000e+03 3.818949999999999748e-05
          2.700000000000000000e+03 3.859300000000000133e-05
          2.800000000000000000e+03 3.834099999999999764e-05
          2.900000000000000000e+03 3.854100000000000250e-05
          3.000000000000000000e+03 3.831750000000000032e-05
          3.100000000000000000e+03 3.826049999999999731e-05
          3.200000000000000000e+03 3.818749999999999987e-05
          3.300000000000000000e+03 3.803199999999999772e-05
          3.400000000000000000e+03 3.841249999999999857e-05
          3.500000000000000000e+03 3.800499999999999950e-05
          3.600000000000000000e+03 3.823450000000000129e-05
          3.700000000000000000e+03 3.801800000000000090e-05
          3.800000000000000000e+03 3.856349999999999763e-05
          3.900000000000000000e+03 3.817250000000000086e-05
          4.000000000000000000e+03 3.845650000000000018e-05
          4.100000000000000000e+03 3.848200000000000188e-05
          4.200000000000000000e+03 3.806049999999999923e-05
          4.300000000000000000e+03 3.848450000000000059e-05
          4.400000000000000000e+03 3.817050000000000325e-05
          4.500000000000000000e+03 3.840299999999999806e-05
          4.600000000000000000e+03 3.807199999999999734e-05
          4.700000000000000000e+03 3.851300000000000209e-05
          4.800000000000000000e+03 3.839200000000000105e-05
          4.900000000000000000e+03 3.794050000000000037e-05
          5.000000000000000000e+03 3.810450000000000084e-05
          5.100000000000000000e+03 3.810899999999999715e-05
          5.200000000000000000e+03 3.800000000000000209e-05
          5.300000000000000000e+03 3.783499999999999944e-05
          5.400000000000000000e+03 3.817950000000000266e-05
          5.500000000000000000e+03 3.782950000000000093e-05
          5.600000000000000000e+03 3.810500000000000193e-05
          5.700000000000000000e+03 3.826000000000000299e-05
          5.800000000000000000e+03 3.787099999999999706e-05
          5.900000000000000000e+03 3.815500000000000315e-05
          6.000000000000000000e+03 3.804099999999999713e-05
          6.100000000000000000e+03 3.819800000000000257e-05
          6.200000000000000000e+03 3.778249999999999952e-05
          6.300000000000000000e+03 3.805949999999999703e-05
          6.400000000000000000e+03 3.786799999999999726e-05
          6.500000000000000000e+03 3.779449999999999872e-05
          6.600000000000000000e+03 3.804600000000000131e-05
          6.700000000000000000e+03 3.775749999999999891e-05
          6.800000000000000000e+03 3.783149999999999854e-05
          6.900000000000000000e+03 3.784200000000000123e-05
          7.000000000000000000e+03 3.812050000000000204e-05
          7.100000000000000000e+03 3.776049999999999871e-05
          7.200000000000000000e+03 3.786100000000000224e-05
          7.300000000000000000e+03 3.800499999999999950e-05
          7.400000000000000000e+03 3.754550000000000162e-05
          7.500000000000000000e+03 3.790799999999999687e-05
          7.600000000000000000e+03 3.769450000000000307e-05
          7.700000000000000000e+03 3.787300000000000144e-05
          7.800000000000000000e+03 3.758550000000000123e-05
          7.900000000000000000e+03 3.787900000000000105e-05
          8.000000000000000000e+03 3.806150000000000142e-05
          8.100000000000000000e+03 3.760549999999999765e-05
          8.200000000000000000e+03 3.828599999999999901e-05
          8.300000000000000000e+03 3.791849999999999957e-05
          8.400000000000000000e+03 3.793399999999999967e-05
          8.500000000000000000e+03 3.764649999999999946e-05
          8.600000000000000000e+03 3.814599999999999697e-05
          8.700000000000000000e+03 3.788699999999999826e-05
          8.800000000000000000e+03 3.815500000000000315e-05
          8.900000000000000000e+03 3.819750000000000147e-05
          9.000000000000000000e+03 3.785599999999999805e-05
          9.100000000000000000e+03 3.832700000000000082e-05
          9.200000000000000000e+03 3.798450000000000199e-05
          9.300000000000000000e+03 3.813300000000000234e-05
          9.400000000000000000e+03 3.789349999999999896e-05
          9.500000000000000000e+03 3.817499999999999957e-05
          9.600000000000000000e+03 3.812799999999999816e-05
          9.700000000000000000e+03 3.786249999999999875e-05
          9.800000000000000000e+03 3.816600000000000016e-05
          9.900000000000000000e+03 3.785750000000000134e-05
          1.000000000000000000e+04 3.810750000000000064e-05
        };
      \addlegendentry{$f$}

      \addplot [thick, c2]
      table [col sep=space] {1.000000000000000000e+00 5.607049999999999789e-05
          1.000000000000000000e+02 6.153649999999999604e-05
          2.000000000000000000e+02 6.153000000000000212e-05
          3.000000000000000000e+02 6.350349999999999940e-05
          4.000000000000000000e+02 6.396399999999999948e-05
          5.000000000000000000e+02 6.452599999999999851e-05
          6.000000000000000000e+02 6.692199999999999386e-05
          7.000000000000000000e+02 6.689400000000000023e-05
          8.000000000000000000e+02 6.832400000000000517e-05
          9.000000000000000000e+02 6.846950000000000573e-05
          1.000000000000000000e+03 6.806799999999999949e-05
          1.100000000000000000e+03 6.895349999999999635e-05
          1.200000000000000000e+03 6.812299999999999812e-05
          1.300000000000000000e+03 6.896899999999999645e-05
          1.400000000000000000e+03 6.843949999999999415e-05
          1.500000000000000000e+03 6.923649999999999347e-05
          1.600000000000000000e+03 6.830799999999999719e-05
          1.700000000000000000e+03 6.885899999999999920e-05
          1.800000000000000000e+03 6.869449999999999764e-05
          1.900000000000000000e+03 6.852999999999999608e-05
          2.000000000000000000e+03 6.821900000000000533e-05
          2.100000000000000000e+03 6.844649999999999595e-05
          2.200000000000000000e+03 6.908550000000000118e-05
          2.300000000000000000e+03 6.882500000000000596e-05
          2.400000000000000000e+03 6.882199999999999939e-05
          2.500000000000000000e+03 6.908949999999999640e-05
          2.600000000000000000e+03 6.865950000000000221e-05
          2.700000000000000000e+03 6.893449999999999534e-05
          2.800000000000000000e+03 6.829249999999999709e-05
          2.900000000000000000e+03 6.877750000000000345e-05
          3.000000000000000000e+03 6.874400000000000454e-05
          3.100000000000000000e+03 6.910300000000000567e-05
          3.200000000000000000e+03 6.830600000000000636e-05
          3.300000000000000000e+03 6.850650000000000554e-05
          3.400000000000000000e+03 6.908300000000000248e-05
          3.500000000000000000e+03 6.876100000000000116e-05
          3.600000000000000000e+03 6.854299999999999748e-05
          3.700000000000000000e+03 6.884050000000000607e-05
          3.800000000000000000e+03 6.908650000000000338e-05
          3.900000000000000000e+03 6.874900000000000195e-05
          4.000000000000000000e+03 6.884800000000000219e-05
          4.100000000000000000e+03 6.904649999999999698e-05
          4.200000000000000000e+03 6.875500000000000155e-05
          4.300000000000000000e+03 6.925850000000000105e-05
          4.400000000000000000e+03 6.812350000000000599e-05
          4.500000000000000000e+03 6.920399999999999675e-05
          4.600000000000000000e+03 6.880350000000000626e-05
          4.700000000000000000e+03 6.919600000000000631e-05
          4.800000000000000000e+03 6.861550000000000060e-05
          4.900000000000000000e+03 6.870700000000000472e-05
          5.000000000000000000e+03 6.897249999999999735e-05
          5.100000000000000000e+03 6.897949999999999915e-05
          5.200000000000000000e+03 6.789350000000000311e-05
          5.300000000000000000e+03 6.863749999999999463e-05
          5.400000000000000000e+03 6.878950000000000266e-05
          5.500000000000000000e+03 6.840550000000000092e-05
          5.600000000000000000e+03 6.862500000000000110e-05
          5.700000000000000000e+03 6.901099999999999368e-05
          5.800000000000000000e+03 6.857299999999999550e-05
          5.900000000000000000e+03 6.884599999999999780e-05
          6.000000000000000000e+03 6.799850000000000295e-05
          6.100000000000000000e+03 6.887550000000000150e-05
          6.200000000000000000e+03 6.841999999999999883e-05
          6.300000000000000000e+03 6.888649999999999851e-05
          6.400000000000000000e+03 6.816050000000000580e-05
          6.500000000000000000e+03 6.834100000000000179e-05
          6.600000000000000000e+03 6.847200000000000443e-05
          6.700000000000000000e+03 6.877250000000000604e-05
          6.800000000000000000e+03 6.788250000000000609e-05
          6.900000000000000000e+03 6.839849999999999912e-05
          7.000000000000000000e+03 6.865100000000000390e-05
          7.100000000000000000e+03 6.839049999999999513e-05
          7.200000000000000000e+03 6.797550000000000673e-05
          7.300000000000000000e+03 6.874750000000000544e-05
          7.400000000000000000e+03 6.783699999999999442e-05
          7.500000000000000000e+03 6.841549999999999574e-05
          7.600000000000000000e+03 6.767199999999999854e-05
          7.700000000000000000e+03 6.857499999999999989e-05
          7.800000000000000000e+03 6.820600000000000393e-05
          7.900000000000000000e+03 6.836350000000000369e-05
          8.000000000000000000e+03 6.915049999999999463e-05
          8.100000000000000000e+03 6.832550000000000168e-05
          8.200000000000000000e+03 6.908099999999999809e-05
          8.300000000000000000e+03 6.882449999999999809e-05
          8.400000000000000000e+03 6.802149999999999917e-05
          8.500000000000000000e+03 6.847749999999999616e-05
          8.600000000000000000e+03 6.893649999999999973e-05
          8.700000000000000000e+03 6.846299999999999825e-05
          8.800000000000000000e+03 6.873049999999999527e-05
          8.900000000000000000e+03 6.885250000000000528e-05
          9.000000000000000000e+03 6.856699999999999590e-05
          9.100000000000000000e+03 6.892800000000000142e-05
          9.200000000000000000e+03 6.813300000000000649e-05
          9.300000000000000000e+03 6.868450000000000282e-05
          9.400000000000000000e+03 6.839299999999999384e-05
          9.500000000000000000e+03 6.922700000000000652e-05
          9.600000000000000000e+03 6.883549999999999511e-05
          9.700000000000000000e+03 6.826350000000000126e-05
          9.800000000000000000e+03 6.890300000000000081e-05
          9.900000000000000000e+03 6.855100000000000147e-05
          1.000000000000000000e+04 6.824200000000000155e-05
        };
      \addlegendentry{$\nabla f$ with source code transformation}

    \end{axis}
  \end{tikzpicture}
  \vspace*{-1.05cm}
  \caption{Wall clock time on Intel Ponte Vecchio GPU of evaluating the function of Fig.~\ref{function_f} and the generated gradient using source code transformation.}
  \label{perf_results_PV}
\end{figure}

The generated gradient code has been tested on an NVIDIA H100 GPU in Fig.~\ref{perf_results_H100}.
As illustrated, the wall-clock time required to evaluate Fig.~\ref{grad_f} is at most twice the wall-clock time of evaluating Fig.~\ref{function_f}.

The performance of the generated gradient has also been evaluated on one of the two AMD Graphics Compute Dies (GCD) of an AMD MI250x as shown in Fig.~\ref{perf_results_mi250x}.
We can observe that the gradient computation takes about twice the wall clock time of evaluating the function.
Finally, the performance of the generated gradient has also been evaluated on Intel Ponte Vecchio GPU as shown in Fig.~\ref{perf_results_PV}.
Kokkos uses a SYCL backend when running on Intel Ponte Vecchio.
This time again, we can observe that the gradient computation takes about twice the wall clock time of evaluating the function.
To the best of our knowledge, this is the first that time a source transformation approach has been used for the automatic differentiation of C++ code on an Intel Ponte Vecchio GPU.

\section{Conclusion}
\label{sec::Conclusion}

Automatic differentiation has became a critical technique with the rise of programming environments such as Jax~\cite{jax2018github,frostig2018compiling}, PyTorch~\cite{paszke2017automatic,NEURIPS2019_9015}, and TensorFlow ~\cite{tensorflow_developers_2022_6574269,tensorflow2015-whitepaper}.
While the Sacado can apply forward mode AD to Kokkos code, no reverse mode AD tools could be used to efficiently differentiate Kokkos parallel code on accelerator architectures.

We demonstrated the feasibility of source transformation-based AD for the automatic generation of performance portable gradient computations for Kokkos-based functions.
We illustrated the performance of the generated code on three GPU architectures.
For these three architectures, the wall-clock time of the automatically generated gradient code is at most $2.17\times$ the wall-clock time of the input function.
To the best of our knowledge, this is the first time that AD has been successfully used for gradient computations of C++ codes on Intel Ponte Vecchio GPUs and therefore the first approach that supports the Aurora supercomputer~\cite{osti_1562918}.

While the results presented in this paper are encouraging, there is still a considerable amount of work that needs to be done in order to apply the proposed approach on complex software such as Albany \cite{Salinger2016,MPASAlbany2018}. We need to develop the differentiable rules of the remaining Kokkos features such as parallel scan, hierarchical parallelism, and SIMD operations. Moreover, we need to tackle the derivative of MPI communication routines. In this paper we only discussed the gradient computations, however applications typically require tangent and Jacobian computations, and possibly Hessian or Hessian-vector-product computations as well. These other derivative computations require different derivation rules and come with other challenges.

There are also some restrictions with our proposed approach that should be resolved to improve the performance of the generated code in reverse mode.
\begin{itemize}
  \item The approach discussed in this paper reuses the thread scheduling of the forward pass for the reverse pass and therefore might introduce unnecessary atomic operations. An optimization that could be implemented is to let the automatic differentiation tool either reschedule the threads for the reverse pass, or to allocate and use shared memory to store temporary computations to minimize the usage of atomic operations.
  \item The number of kernels of the generated gradient source code is multiplied by two compared to the number of kernels of the input source code. The number of synchronization points between the host and device have therefore increased as well. A possible approach to mitigate that effect is to automatically detect possibilities for kernel fusion.
  \item As illustrated in the example of section~\ref{sec::Results}, our current approach does not yet support taping of nonlinear operations within kernels.
        A way to efficiently implement that taping would be to automatically deduce the sizes of the memory required to store all the recorded data and to allocate that memory once outside the kernel.
\end{itemize}

We will investigate these ideas mentioned above in the future.

\section{Acknowledgements}
This work was supported by the S4PST project, and the Laboratory Directed Research and Development program at Sandia National Laboratories. Sandia National Laboratories is a multimission laboratory managed and operated by National Technology \& Engineering Solutions of Sandia, LLC, a wholly owned subsidiary of Honeywell International Inc., for the U.S. Department of Energy’s National Nuclear Security Administration under contract DE-NA0003525.
This work is partially supported by the National Science Foundation under Grant OAC-2311471.
This paper describes objective technical results and analysis. Any subjective views or opinions that might be expressed in the paper do not necessarily represent the views of the U.S. Department of Energy or the United States Government. SAND2025-08673O.

\bibliographystyle{siam}
\bibliography{reference}

\end{document}